# Spectral Properties and Energy Injection in Mercury's Magnetotail Current Sheet

Xinmin Li[1], Chuanfei Dong[1], Liang Wang[1], Sae Aizawa[2], Lina Z. Hadid[2], Chi Zhang[1], Hongyang Zhou[1], James A. Slavin[3], Jiawei Gao[1], Mirko Stumpo[4] and Wei Zhang[1]

1. *Center for Space Physics, Boston University, Boston, MA 02215, USA*

2. *Laboratoire de Physique des Plasmas (LPP), CNRS, Observatoire de Paris, Sorbonne Université, Université Paris Saclay, Ecole polytechnique, Institut Polytechnique de Paris, Palaiseau 91120, France*

3. *Department of Climate and Space Sciences and Engineering, University of Michigan, Ann Arbor, MI 48109, USA*

4. *INAF-Institute for Astrophysics and Space Planetology, Via del Fosso del Cavaliere, 00133 Roma, Italy*

Corresponding author: X. Li (xli8@bu.edu) and C. Dong (dcfy@bu.edu)

**Key Points**

- Most magnetotail current sheets at Mercury are turbulent, with a spectral breaks between inertial and kinetic ranges
- Spectral slopes show dawn-dusk asymmetry, with shallower inertial and steeper kinetic slopes on the dawn side
- Transverse magnetic components display shallow slopes near −1 in the inertial range, suggesting that energy is injected at ion scales

**Abstract**

Mercury's magnetotail hosts a thin and highly dynamic current sheet (CS), where magnetic reconnection and strong fluctuations frequently occur. Here, we statistically analyze magnetic field power spectra across 370 magnetotail CSs observed by MESSENGER. About 20% of the events are quasi-laminar, showing single power-law spectra, whereas ~80% are turbulent, exhibiting a spectral break separating inertial and kinetic ranges. A dawn–dusk asymmetry is identified: inertial-range slopes are systematically shallower on the dawnside, whereas kinetic-range slopes are steeper, indicating more developed turbulence there, consistent with the higher occurrence of reconnection-related processes on the dawnside. Component analysis shows that the transverse components, orthogonal to the tail-aligned principal field ($B_X$), display shallow slopes near −1 in the inertial range, suggesting energy injection at ion scales rather than a classical inertial range. These results demonstrate that Mercury's unique plasma environment fundamentally reshapes the initiation of turbulence and the redistribution of energy in the magnetotail.

**Plain Language Summary**

Mercury's magnetotail contains a thin and rapidly changing current sheet where magnetic reconnection and plasma turbulence frequently occur. Using data from the MESSENGER spacecraft, we analyzed 370 crossings of this current sheet to understand how magnetic fluctuations behave at different scales. We find that most current sheets are turbulent and show a clear transition between large-scale (inertial) and small-scale (kinetic) fluctuations, while a smaller fraction remain smooth and laminar. The turbulence also differs between the dawn and dusk sides of the tail, with stronger and steeper fluctuations on the dawn side. In addition, magnetic variations perpendicular to the main tail field show unusually flat spectra, suggesting that energy is injected directly at ion scales rather than cascading from larger scales. These results reveal that Mercury's fast magnetic cycle and extreme space environment fundamentally change how turbulence begins and how energy is distributed in its magnetotail.

## 1. Introduction

Mercury hosts the smallest and most solar-wind–driven magnetosphere in the solar system. Compared with Earth, its close proximity to the Sun, weak intrinsic magnetic field, and compressed magnetopause create a system that is far more dynamic [*Aizawa et al.*, 2023; *Anderson et al.*, 2011; *Diego et al.*, 2020; *Dong et al.*, 2019; *Guo et al.*, 2023; *Q Lu et al.*, 2022; *Milillo et al.*, 2020; *Orsini et al.*, 2022; *J Slavin et al.*, 2019; *J A Slavin*, 2004; *R Wang et al.*, 2024], with magnetic flux circulating through a Dungey cycle process on timescales of only a few minutes [*S M Imber and Slavin*, 2017; *J A Slavin et al.*, 2009; *W J Sun et al.*, 2015], which is much shorter than ~1 hour at Earth [*Baker et al.*, 1996]. A direct consequence of this rapid process is the formation of a highly stretched and dynamic magnetotail, in which the cross-tail current sheet plays a central role in magnetic energy conversion and particle acceleration [*Gershman et al.*, 2014; *Shi et al.*, 2022; *J A Slavin et al.*, 2010; *J A Slavin et al.*, 2012; *W Sun et al.*, 2017; *W Sun et al.*, 2018; *Zhao et al.*, 2020].

Previous observations indicate that Mercury's magnetotail current sheet is remarkably thin, with a typical thickness of about 0.4 $R_M$ (where $R_M$ = 2,440 km, denotes Mercury's radius), corresponding to several proton inertial lengths [*Dewey et al.*, 2020; *Gershman et al.*, 2014; *Poh et al.*, 2017b; *Rong et al.*, 2018]. Such ultrathin current sheets are inherently favorable for magnetic reconnection. Observations from the MErcury Surface, Space ENvironment, GEochemistry, and Ranging (MESSENGER) spacecraft have revealed frequent reconnection signatures and associated structures, including flux ropes and dipolarization fronts, within Mercury's magnetotail [*DiBraccio et al.*, 2015; *J A Slavin et al.*, 2012; *W Sun et al.*, 2016; *Sundberg et al.*, 2012; *Zhong et al.*, 2020; *Zhong et al.*, 2018]. Similar to Earth [*J A Slavin et al.*, 1985; *Z Zhang et al.*, 2024],

Mercury's cross-tail current sheet exhibits a dawn–dusk asymmetry in its structure, with the duskside generally more stretched, resulting in smaller normal $B_Z$ and a thinner current sheet compared with the dawnside [*Poh et al.*, 2017a; *Rong et al.*, 2018]. However, unlike Earth, where reconnection is often favored on the duskside [*S Imber et al.*, 2011; *J Liu et al.*, 2013; *Z Liu et al.*, 2025; *S Lu et al.*, 2016], at Mercury it occurs preferentially on the dawnside [*Y Chen et al.*, 2019; *Poh et al.*, 2017a; *Smith et al.*, 2017; *W Sun et al.*, 2016]. This contrast may be related to enhanced heavy planetary ion populations on the duskside [*Y Chen et al.*, 2024], which increase plasma inertia and suppress reconnection onset, despite the current sheet there being thinner and more stretched [*Poh et al.*, 2017a]. Alternatively, simulations show that once reconnection initiates, the X-line and associated outflows tend to spread preferentially toward the dawnside in Mercury's magnetotail [*Y Chen et al.*, 2019; *Y H Liu et al.*, 2019], providing an additional mechanism that may account for the observed dawn–dusk asymmetry.

Owing to the highly dynamic activity in Mercury's magnetotail, the magnetotail current sheet often hosts strong magnetic field fluctuations, providing a natural environment for plasma turbulence to develop. Although many studies have examined Mercury's magnetotail structure and reconnection phenomena, the properties of these magnetic fluctuations (and their turbulent characteristics) remain largely unexplored. The turbulence within the terrestrial magnetotail current sheet has been suggested to enhance energy conversion and particle acceleration processes [*Ergun et al.*, 2018; *Fu et al.*, 2017; *S Y Huang et al.*, 2022; *Li et al.*, 2022a; *Li et al.*, 2022b]. Therefore, understanding the magnetic fluctuation in Mercury's current sheet is crucial for evaluating how energy conversion processes operate in Mercury's extreme magnetospheric environment.

In this paper, we use magnetic field data from MESSENGER to statistically characterize turbulence in Mercury's cross-tail current sheet. Specifically, we investigate the power spectral densities (PSDs) of the magnetic field and quantify the fluctuation spectral slopes across both inertial and kinetic ranges. We present the first statistical measurements of spectral slopes in Mercury's magnetotail current sheet, including their dawn–dusk asymmetries, and discuss the implications for turbulence and energy conversion in Mercury's unique plasma environment.

## 2. Data and Methods

We use 20 Hz magnetic field measurements from the Magnetometer (MAG) [*Anderson et al.*, 2007] and Fast Imaging Plasma Spectrometer (FIPS) [*Andrews et al.*, 2007] onboard the MESSENGER spacecraft during its entire orbital phase around Mercury from 2011 to 2015. The data are analyzed in the Mercury Solar Magnetospheric (MSM) coordinate system, where +X axis points from the offset location of Mercury's dipole toward the Sun, the +Y axis lies in Mercury's equatorial plane and it is positive in the direction opposite to planetary motion, and +Z completes the right-handed orthogonal system. This study focuses on current sheet crossings in the magnetotail, within $-0.5$ $R_M < X < -4$ $R_M$, where $R_M$ is Mercury's radius [*Poh et al.*, 2017a; *Rong et al.*, 2018].

A total of 370 magnetotail current sheet events are identified based on the following criteria: (1) A well-defined overall reversal of $B_X$, with small-scale perturbations suppressed by applying a 1 minute moving average to $B_X$ during reversal detection. (2) The neutral sheet ($B_X=0$) is located within $-1$ $R_M < Y < 1$ $R_M$, which helps exclude magnetopause current sheets in the flank region. (3) A well-defined transition into the relatively steady lobe magnetic fields on either side of the current sheet following the reversal. Each event is centered on the $B_X$ reversal and spans a 20-minute interval. This 20-minute interval is chosen empirically to encompass the full current sheet crossing while minimizing contamination from adjacent regions. If multiple consecutive reversals occur within a single orbit, the temporally central $B_X$ reversal is chosen as the event center. We exclude events in which the spacecraft leaves the magnetotail current sheet during this period, as indicated by a significant increase in magnetic field strength.

For each crossing, we compute the PSD of the magnetic field using the fast Fourier transform (FFT) over the 20-minute interval. To identify spectral breaks and obtain spectral slopes, we implement a fitting routine that compares single- and double-power-law models in log–log space. The PSD is first computed and log-transformed along with the frequency axis. We exclude data points that are strongly deviated by narrowband fluctuations, and restrict the fitting to the frequency range from 3 mHz to 5 Hz, which covers both MHD and sub-ion scales separated by the spectral break near the proton cyclotron frequency (with an average $f_{cp}$ of approximately 0.2 Hz in this study). A single-slope linear fit is applied over the entire frequency range, and its residual sum of squares (RSS) is calculated. To test for a break, we then sequentially shift a hypothetical break point across the data points. For each assumed break location, the spectrum is divided into two segments, both fitted with straight lines, and the combined RSS is calculated. The break frequency is taken from the position that minimizes the RSS. A double-slope model is adopted only if it reduces the RSS by more than ~30% compared to the single-slope fit; otherwise, a single power law is retained. The 30% criterion is introduced to avoid spurious breaks, and varying this threshold within 15–50% does not qualitatively affect our results.

## 3. Examples and Statistical study

### 3.1 Representative Cases

We classify the 370 current sheets into two categories based on the morphology of their magnetic power spectra. Type 1 events are characterized by spectra that can be described by a single power law over the analyzed frequency range, whereas Type 2 events display a distinct spectral break, with different slopes on either side of the break frequency. Unless otherwise specified, spectra shown in this paper refer to the trace of the power spectrum of the total magnetic field, obtained as the sum of the power in the three magnetic field components.

Figure 1 presents the representative examples of each type. The Type 1 event (left column) is observed by MESSENGER from 07:04 UT to 07:24 UT on 6 November 2013, and the Type 2 event is observed (right column) from 21:18 UT to 21:38 UT on 22 November 2011. In both cases, MESSENGER traverses the current sheet from the southern side to the northern side, as indicated by the reversal of $B_X$ from negative to positive (blue traces in Figures 1c and 1g). Enhanced fluxes of high-energy (>1 keV) protons are observed within the current sheet in both events (Figures 1a and 1e), consistent with previous observations of magnetotail current sheets at both Mercury and Earth [*Gershman et al.*, 2014; *Li et al.*, 2019; *W Sun et al.*, 2017; *R Wang et al.*, 2018; *R S Wang et al.*, 2020].

In the Type 1 event, the magnetic field magnitude and three components (Figures 1b and 1c) vary smoothly across the current sheet, with minimal magnetic perturbations superimposed on the larger-scale change. Its power spectrum is well described by a single power law, with a slope of -2.0 (Figure 1d). This example represents a quasi-laminar current sheet, that maintains a smooth and coherent large-scale magnetic structure with little to no evidence of magnetic fluctuations. In contrast, the Type 2 event shows pronounced fluctuations in the magnetic field (Figures 1f and 1g), particularly near the center of the current sheet. The corresponding power spectrum displays a break at around 0.3 Hz (about 0.5 $f_{cp}$), separating an inertial-range slope of -1.66 from a steeper kinetic-range slope of -2.30. These two examples illustrate the contrasting spectral signatures used for classification, which are further quantified in the statistical distribution presented in the following section. It worth to note that such enhanced central fluctuations reflect internal inhomogeneity of the current sheet, the detailed spatial evolution of which will be examined in future studies.

### 4.2 Statistical Distribution of Spectral Slopes

Figure 2 summarizes the statistical distributions of spectral slopes for all 370 current sheet crossings. Type 1 events (Figure 2a) account for 20.5% of the sample and are described by a single spectral slope. Their slopes range from -1.6 to -2.3, with the majority clustering narrowly around −2.0, consistent with the expected slope for an idealized Harris-type current sheet (see the Supplementary Material).

The remaining 79.5% of events are classified as Type 2 (Figure 2b), characterized by two distinct slopes separated by a spectral break ($f_b$). In the inertial range ($f < f_b$; red bars in Figure 2b), the slopes span approximately -1.1 to -2.2, peaking near −5/3, the value consistent with a Kolmogorov-like spectrum [*Kolmogoroff*, 1941]. In the kinetic range ($f > f_b$; blue bars in Figure 2b), slopes range from about -1.9 to -2.6, with most values clustering around -2.2. The slope distributions in the two ranges have only a small degree of overlap, and show distinct central tendencies, indicating that the slopes steepen from the inertial-range to the kinetic-range. Building on this statistical baseline, we then investigate how these spectral characteristics differ between the dawn and dusk sides of the magnetotail.

## 4.3 Dawn–Dusk Asymmetry of Spectral Slopes

Since Type 1 events are relatively few in number (20.5% of the sample) and have spectral slopes concentrated within a narrow range around -2.0, the dawn–dusk comparison is performed only for Type 2 events, which show sufficient variability in both inertial-range and kinetic-range slopes. Here, dawnside refers to events with Y<0 in MSM coordinates, and duskside refers to events with Y>0. Figure 3 compares the spectral slope distributions between the dawn and dusk sides of the magnetotail for Type 2 events.

In the inertial range ($f<f_b$; Figure 3a), slopes on both the dawn (purple bars) and dusk sides (blue bars) fall primarily between −2.2 and −1.2, with central tendencies near −5/3. Although the two distributions broadly overlap, their detailed shapes differ. The dusk-side distribution is relatively symmetric with a single prominent peak, while the dawn-side distribution is slightly broader and exhibits a secondary peak around −1.45, suggesting a greater occurrence of shallower slopes (the spectral power decreases more slowly with increasing frequency). In kinetic range ($f>f_b$; Figure 3b), slopes on the dusk side (green bars) are concentrated between −2.35 and −1.95, with a clear and symmetric peak around −2.2. However, the dawn-side slopes extend preferentially toward steeper values (the spectral power decreases more fastly with increasing frequency) and exhibit a secondary peak near −2.45, indicating an enhanced occurrence of steeper kinetic-range slopes.

The histogram results above (Figures 3a and 3b) demonstrate a noticeable dawn-dusk asymmetry in both the inertial range and kinetic range. To further investigate how this asymmetry varies with downtail distance, we examine the spatial distribution of median slopes in the X-Y plane in the MSM coordinate system. To ensure the reliability of median values, relatively large bin sizes (0.4 $R_M$, where $R_M$ is Mercury's radius) are used, and only bins containing at least three events are included. As a result, the X range in the maps is restricted to -3.0 $R_M$ to -1.2 $R_M$. The dawn–dusk asymmetry becomes more visible in the 2D map (Figures 3c-3d). And this dawn–dusk asymmetry appears to increase with downtail distance.

### 4.4 Power Spectral Density of Individual Magnetic Field Components

In the following section, we examine the spectral slopes of the three magnetic field components separately in the MSM coordinate system (Figure 4). For $B_X$, most events (about 60%) follow a single power law with slopes concentrated near -2.0 (green bars in Figure 4a). For other events with a spectral break, the inertial-range slopes cluster around -1.8 (blue bars in Figure 4a) and the kinetic-range slopes around -2.15 (red bars in Figure 4a), both lying closer to -2.0 compared with the corresponding slopes of the total magnetic field described earlier. In both the inertial and kinetic ranges, the dusk-side and dawn-side distributions nearly overlap (Figures 4b and 4c), suggesting no pronounced dawn-dusk asymmetry in the $B_X$ component.

For $B_Y$ and $B_Z$, only a small fraction of events follows a single power law, about 5.4% for $B_Y$ (green bars in Figure 4d) and 18.1% for $B_Z$ (green bars in Figure 4g). The slopes of these events also cluster near -2.0. For other events with a spectral break, the inertial-range slopes are markedly shallower than those of $B_X$, with both $B_Y$ and $B_Z$ peaking around -1.2 (red bars in Figures 4d and 4g). In the kinetic range, the slopes for both $B_Y$ and $B_Z$ are steeper than $B_X$, concentrating near -2.4 (blue bars in Figures 4d and 4g). The dawn-dusk asymmetry is evident in both $B_Y$ and $B_Z$ (Figures 4e, 4f, 4h and 4i), consistent with total magnetic field results. This dawn–dusk asymmetry is also evident in the 2D slope maps of $B_Y$ and $B_Z$ (Figure S1 in Supplementary Material).

### 5. Discussions and Conclusion

This study presents the first statistical characterization of magnetic spectra within Mercury's magnetotail current sheet. Unlike previous spectral studies performed in the solar wind or magnetosheath [*C H K Chen and Boldyrev*, 2017; *L Z Hadid et al.*, 2018; *SY Huang et al.*, 2020; *Sahraoui et al.*, 2009], where the background field is typically homogeneous, our analysis focuses on the magnetotail current sheet, a coherent structure with strong magnetic shear. In such a system, the current sheet can be approximated by a Harris-type configuration, where the dominant magnetic component ($B_X$) provides a broadband background with a characteristic spectral slope near −2.0 (Figure S2 in Supplementary Material).

Our results show that a minority of events (Type 1; ~20%) display very small magnetic fluctuations, retain a largely laminar Harris-type structure, and exhibit single power-law spectra with slopes clustered around −2.0, reflecting the spectral signature of a coherent current sheet with weak turbulence. By contrast, the majority (Type 2; ~80%) of events develop strong fluctuations and evolve into a turbulent state. In general, turbulence is expected to produce characteristic slopes of about −5/3 (shallower than -2.0) in the inertial range, and −2.3 to −3.1 (steeper than -2.0) in the kinetic range [*Alexandrova et al.*, 2008; *Boldyrev and Perez*, 2009; *Goldstein et al.*, 1994; *ShiYong Huang et al.*, 2014]. Thus, once turbulence develops in the current sheet, the spectra depart from the Harris-type baseline: the inertial range becomes systematically shallower than −2.0, while the kinetic range becomes steeper.

The component analysis further supports this interpretation. In Mercury's magnetotail, because the background magnetic field of the current sheet is primarily carried by the $B_X$ component, its spectra remain strongly tied to the coherent current sheet structure, with the majority (~60%) described by a single power law near −2.0. This fraction is much higher than for $B_Y$ (5%) and $B_Z$ (18%). Even when spectral breaks occur, $B_X$ slopes in both the inertial and kinetic ranges remain closer to −2.0 than those of $B_Y$ and $B_Z$. Moreover, a notable fraction of $B_Z$ spectra (about 18%) follow a single power-law distribution. This may reflect two effects: (1) current sheet flapping [*Poh et al.*, 2017b; *C Zhang et al.*, 2020], which tilts the sheet and projects part of the $B_X$ component into Z direction, and (2) $B_Z$ is affected by the Mercury's dipole field (Figure S3 in Supplementary Material), which introduces a smooth background variation during current sheet crossings.

Moreover, the component analysis shows that the low-frequency slopes of the $B_Y$ and $B_Z$ components (referred to as the inertial range in this paper) are often much shallower, sometimes approaching −1, far from the expected values (−5/3). Such shallow slopes likely correspond to an energy-injection regime rather than a fully developed inertial range energy cascade. Similar behavior has been reported in planet magnetosheath near the bow shock, where the downstream turbulence does not have sufficient time to develop a classical inertial range [*L Hadid et al.*, 2015; *SY Huang et al.*, 2017; *SY Huang et al.*, 2020]. By analogy, Mercury's extremely short Dungey cycle [*S M Imber and Slavin*, 2017; *J A Slavin et al.*, 2009] likely leaves insufficient time for turbulence in the magnetotail current sheet to fully develop, which limits the formation of a well-developed inertial range. Such energy injection may be associated with local processes inside the current sheet, such as magnetic reconnection or other plasma instabilities, which can directly inject energy into sub-ion-scale fluctuations. Thus, turbulence may initially develop at kinetic scales and then undergo an inverse cascade toward larger scales, but the process is likely limited by Mercury's rapid Dungey cycle.

Another key feature revealed by our analysis is the dawn–dusk asymmetry of the spectral slopes. In the inertial range, dawn-side slopes are systematically shallower than those on the duskside, whereas in the kinetic range they become steeper. This pattern is consistent across both the total magnetic field and the $B_Y$ and $B_Z$ components, suggesting that turbulence is stronger on the dawnside, disrupting the coherent current sheet structure and driving the slopes further away from −2 than on the dusk side. Moreover, although turbulence CSs dominate the overall current sheet population, laminar CSs occur more frequently on the dusk side than the dawn side (Figure S4 in the Supplementary Material). Such asymmetry is likely linked to reconnection-related activity, as previous studies have shown that reconnection and associated structures, such as flux ropes and dipolarization fronts, occur preferentially on the dawn side, thereby enhancing magnetic fluctuations and making the dawn-side current sheet more turbulent [*Dewey et al.*, 2020; *Poh et al.*, 2017a; *W Sun et al.*, 2016]. Therefore, turbulent (Type-2) current sheets likely represent reconnection-active configurations characterized by strong gradients, multiscale structures, and enhanced magnetic fluctuations.

We also examined the thickness of the magnetotail current sheet using the method of *Poh et al.* [2017b]. The results show a clear dawn–dusk asymmetry, with statistically thicker current sheets on the dawnside (Supplementary Figure S5). However, the thickness exhibits a broad distribution at each cross-tail position, and thin current sheets are present on both sides. Further analysis indicates that thinner current sheets tend to exhibit more turbulent spectral characteristics. These results suggest that, while the dawnside magnetotail is globally more turbulent, turbulence and reconnection locally preferentially develop in thinner current sheets.

In summary, this study presents the first comprehensive statistical characterization of magnetic spectral slopes in Mercury's magnetotail current sheet. We show that most current sheets evolve into a turbulent state, with spectra departing from the Harris-type baseline and exhibiting dawn–dusk asymmetry, with more developed turbulence on the dawnside. A central finding is that the unusually shallow low-frequency slopes of the $B_Y$ and $B_Z$ components are consistent with ongoing energy injection at ion scales rather than a well-developed inertial range energy cascade. These results demonstrate that Mercury's unique plasma environment fundamentally reshapes the processes of energy injection and redistribution in planetary magnetotails, and they highlight the broader importance of cross-planetary comparisons for establishing a universal framework of turbulence, reconnection, and multiscale energy conversion in space plasmas.

**Acknowledgments**

This work is supported by NASA grant 80NSSC23K0894, DOE grant DE-SC0024639, and Alfred P. Sloan Research Fellowship. We thank the entire MESSENGER team and instrument principal investigators for providing and calibrating data.

**Open Research**

The MESSENGER data used here are available from the Planetary Data System. The magnetic field data can be accessed at https://pds-ppi.igpp.ucla.edu/collection/urn:nasa:pds:mess-mag-calibrated:data-mso, and the plasma data are available at https://pds-ppi.igpp.ucla.edu/collection/urn:nasa:pds:mess-epps-fips-derived:data-espec. The data have been analyzed, and plotted using the SPEDAS software (Version 6.0) [*Angelopoulos et al.*, 2019], which can be downloaded via the Downloads and Installation page (http://spedas.org/blog/).

**Conflict-of-interest statement**

The authors have no conflicts of interest to disclose.

# Figures and Figure Captions

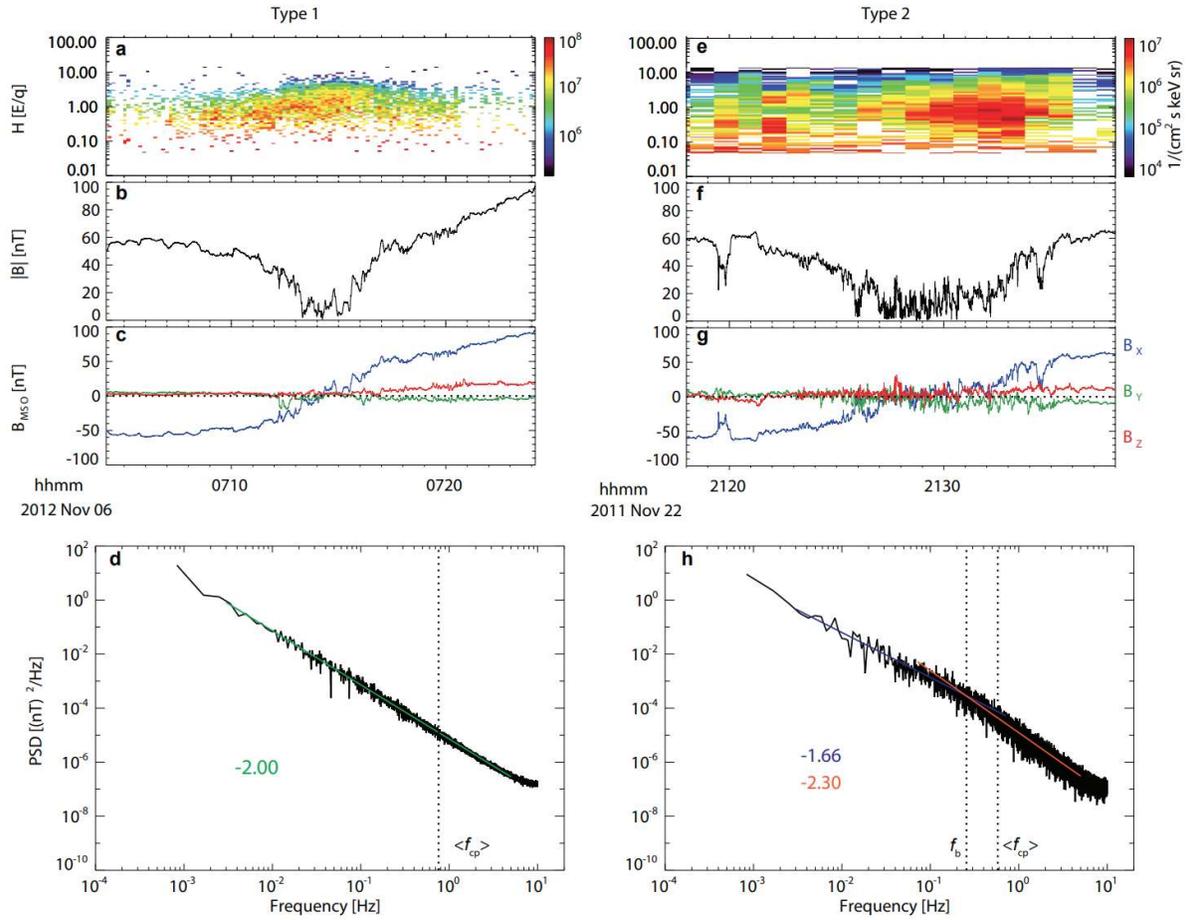

**Figure 1. Two representative current sheet crossings observed by MESSENGER.** Panels (a–d) show a quasi-laminar event on 2012 November 6: (a) ion energy spectrogram; (b) magnetic field magnitude; (c) magnetic field components in the MSM coordinate system; and (d) magnetic field power spectral density (PSD) fitted with a single power law (slope ≈ −2.00). Panels (e–h) show a turbulent event on 2011 November 22, with corresponding PSD exhibiting a spectral break and two fitted slopes of −1.66 and −2.30 in the inertial and kinetic-ranges, respectively. The dashed lines in (d) and (h) represent the average proton cyclotron frequency.

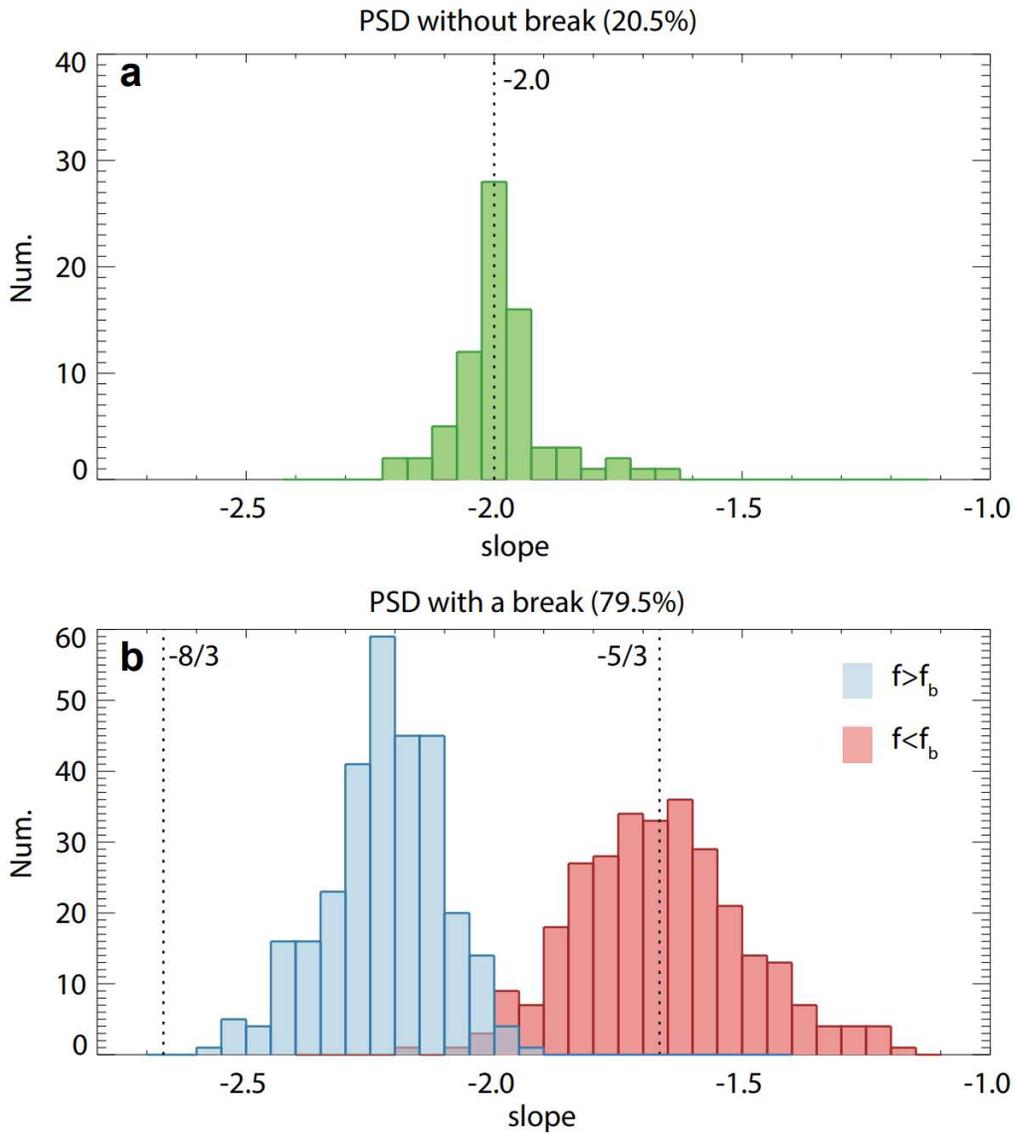

**Figure 2. Statistical distributions of magnetic spectral slopes**. (a) Histogram of slopes for current sheets without spectral breaks (20.5% of the sample). (b) Histogram of inertial-range (red) and kinetic-range (blue) slopes for current sheets with breaks (79.5%). Vertical dashed lines mark theoretical predictions (e.g., −5/3, −8/3).

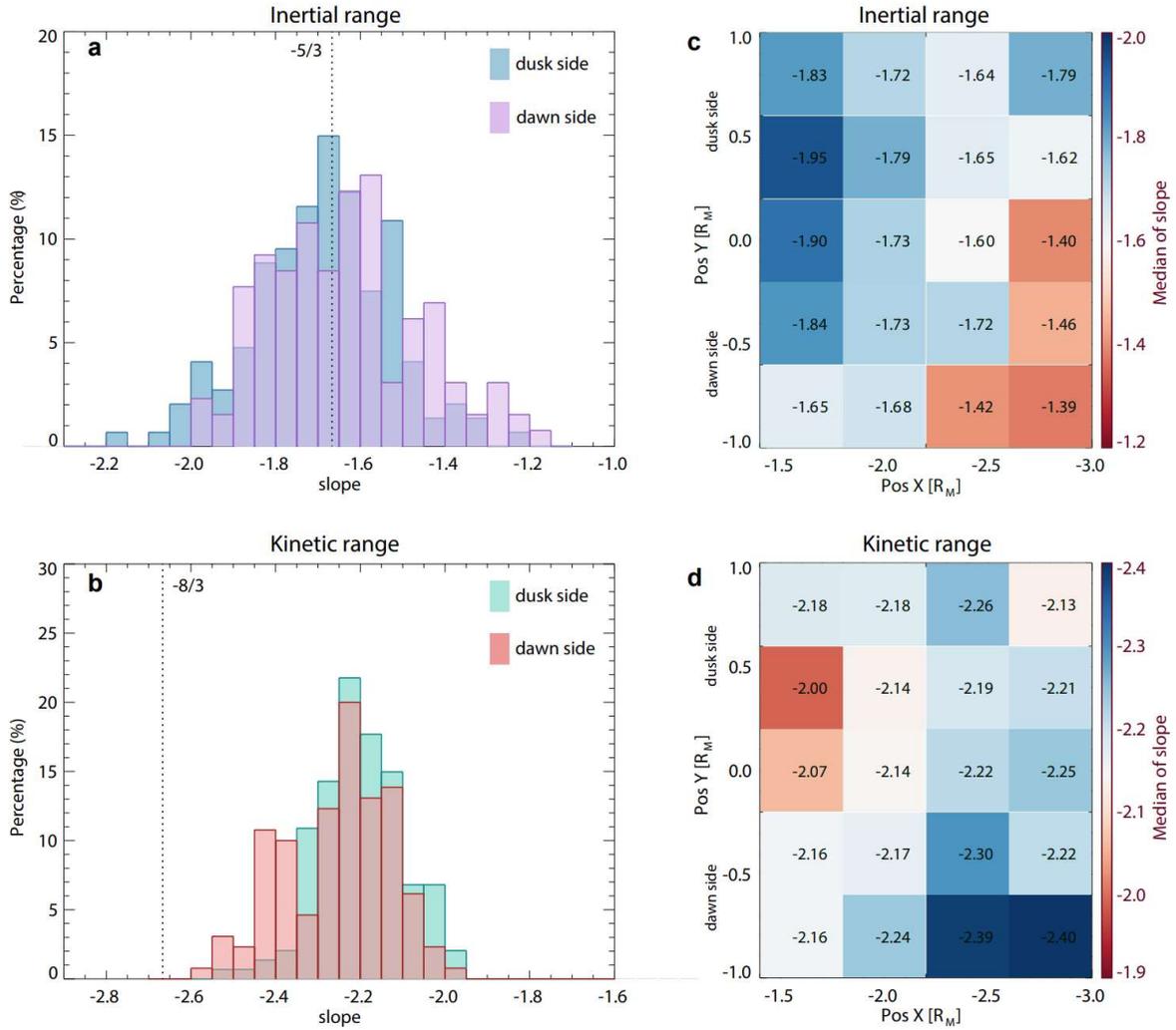

**Figure 3. Dawn–dusk asymmetry in magnetic spectral slopes.** (a) and (b) Histograms of inertial-range (a) and kinetic-range (b) slopes for events on the dawn (purple/red) and dusk (blue/green) sides. (c) and (d) 2D maps of the median slope in the X–Y plane in MSM coordinates, showing spatial variation across the magnetotail in the inertial (c) and kinetic (d) ranges.

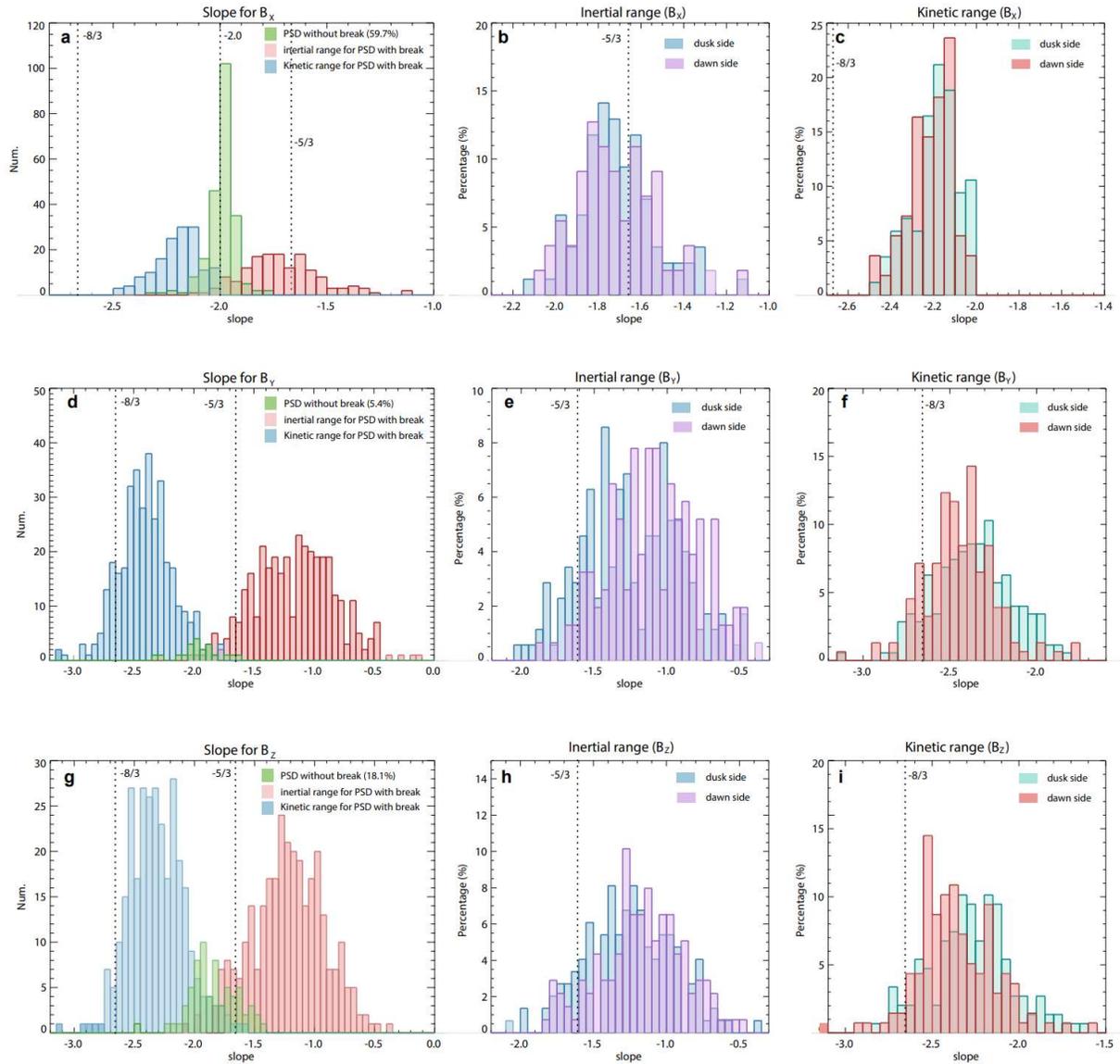

**Figure 4. Magnetic spectral slope distributions for individual field components.** (a) shows the histogram of spectral slopes for $B_X$: green bars indicate slopes for events without spectral breaks, while red and blue bars represent inertial-range and kinetic-range slopes, respectively, for events with breaks. (b) - (c) show the dawn–dusk asymmetry for $B_X$ in the inertial- and kinetic-range slopes, respectively, for events with spectral breaks. Panels (d)–(f) and (g)–(i) follow the same format as (a)–(c), but for the $B_Y$ and $B_Z$ components, respectively.

Supporting Information for

**Spectral Properties and Energy Injection in Mercury's Magnetotail Current Sheet**

Xinmin Li[1], Chuanfei Dong[1], Liang Wang[1], Sae Aizawa[2], Lina Z. Hadid[2], Chi Zhang[1], Hongyang Zhou[1], James A. Slavin[3], Jiawei Gao[1], Mirko Stumpo[4] and Wei Zhang[1]

1. *Center for Space Physics, Boston University, Boston, MA 02215, USA*

2. *Laboratoire de Physique des Plasmas (LPP), CNRS, Observatoire de Paris, Sorbonne Université, Université Paris Saclay, Ecole polytechnique, Institut Polytechnique de Paris, Palaiseau 91120, France*

3. *Department of Climate and Space Sciences and Engineering, University of Michigan, Ann Arbor, MI 48109, USA*

4. *INAF-Institute for Astrophysics and Space Planetology, Via del Fosso del Cavaliere, 00133 Roma, Italy*

Corresponding author: X. Li (xli8@bu.edu) and C. Dong (dcfy@bu.edu)

**Contents of this file**

Figure S1-S5

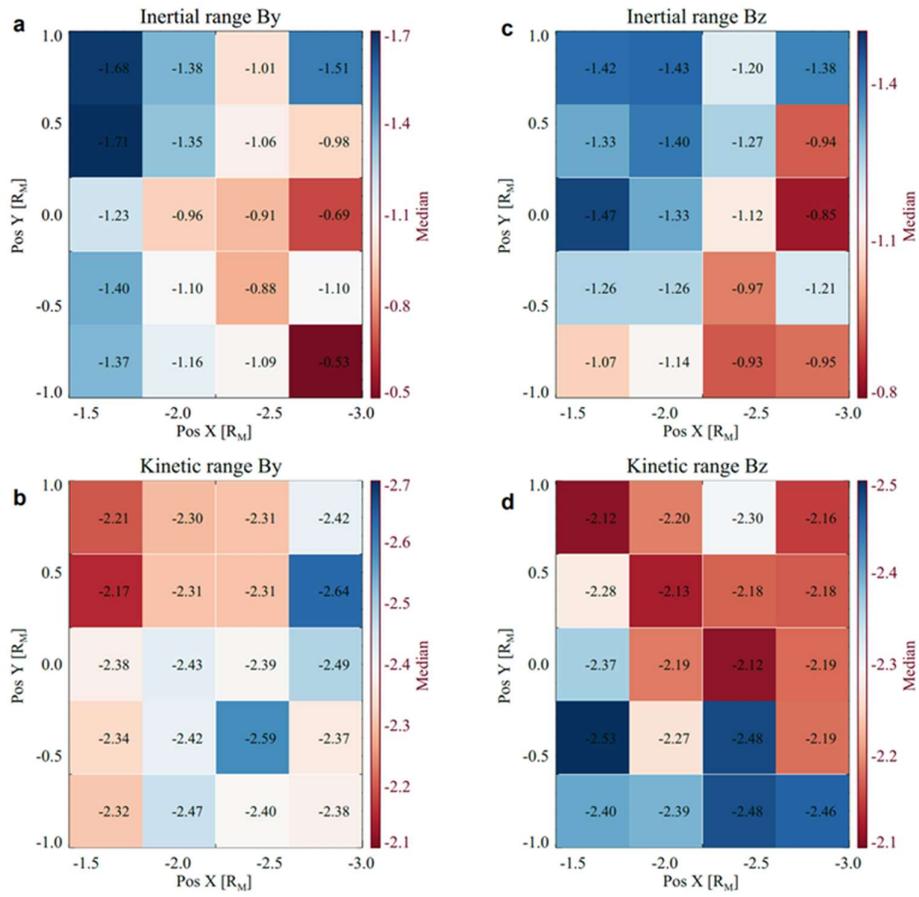

Figure S2. Two-dimensional maps of the median spectral slope in the X–Y plane in Mercury Solar Magnetospheric (MSM) coordinates, showing the spatial variation of $B_Y$ and $B_Z$ across the magnetotail in both the inertial and kinetic ranges. These maps confirm the clear dawn–dusk asymmetry.

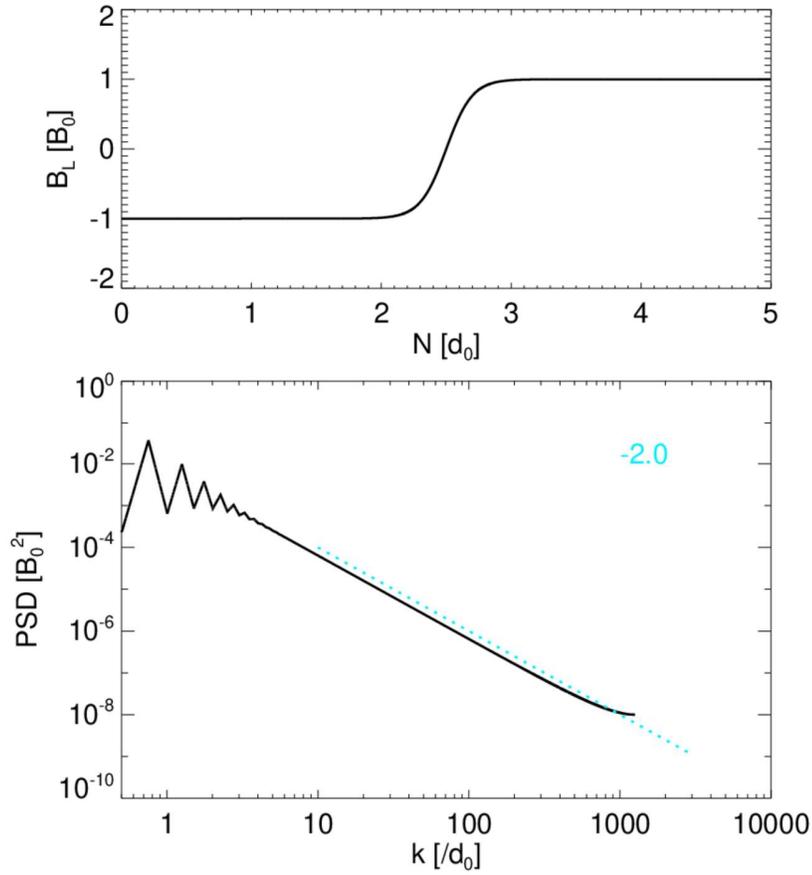

**Figure S2. *FFT of a Harris current sheet.*** Top: magnetic field profile $B(N) = B_0 \tanh(\frac{N}{d_0})$. Bottom: power spectral density P(k). At larger k, the spectrum follows a $k^{-2}$ scaling (slope $-2$).

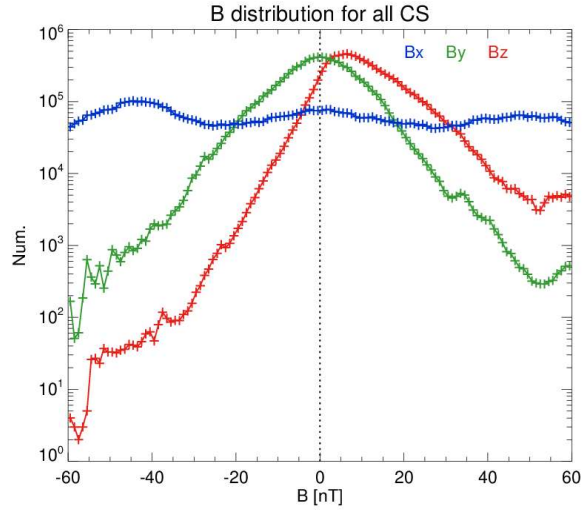

Figure S3. Distributions of three components of magnetic field within the 370 current sheets. The distributions of Bx and By are nearly symmetric about zero, whereas Bz exhibits a clear excess of positive values. This systematic asymmetry indicates that the measured Bz component is influenced by Mercury's northward dipole field.

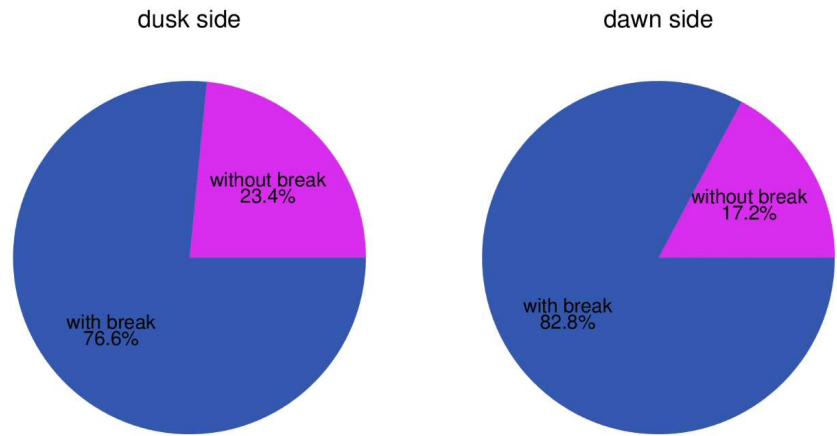

Figure S4. Pie charts showing the fractions of events with (blue; type II) and without (magenta; Type I) spectral breaks on the dusk (left) and dawn (right) sides. Even though Type II events dominate the current sheet population, Type I events occur more frequently on the duskside, whereas Type II events preferentially occur on the dawnside. This distribution is consistent with our conclusion that current sheets on the dawnside are generally more turbulent, leading to a higher proportion of Type II (turbulent) events there.

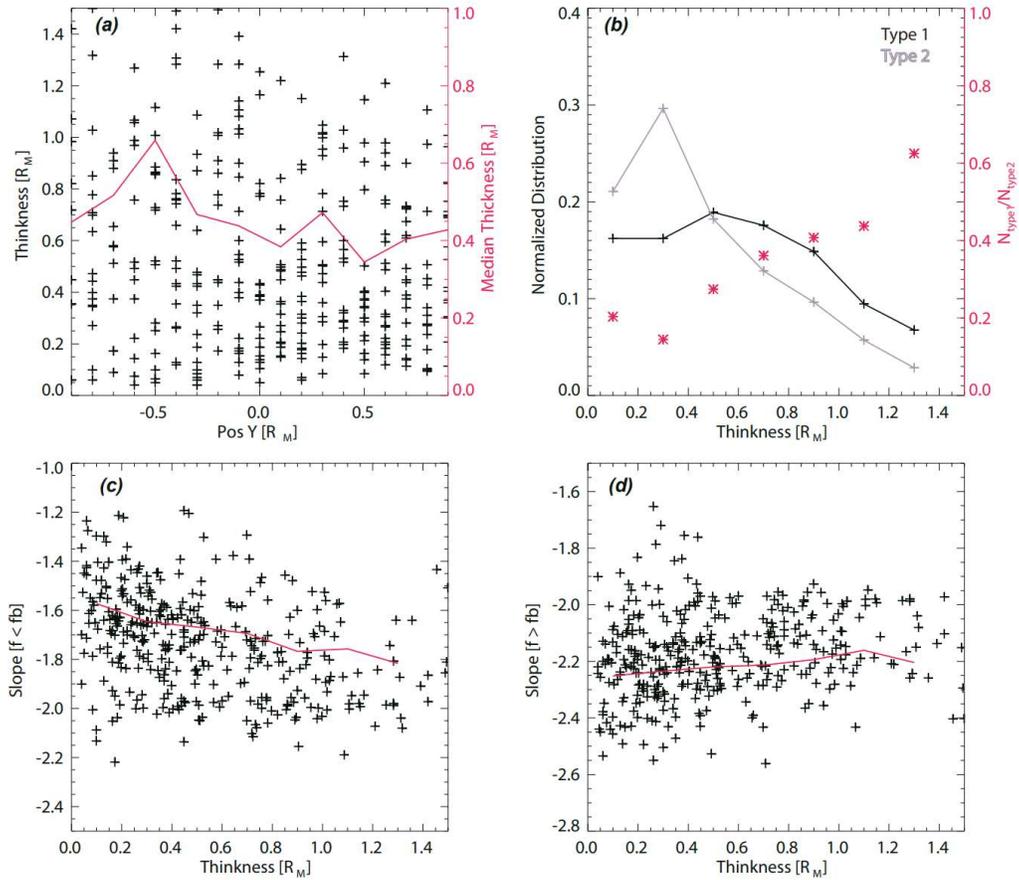

Figure S5. (a) Current sheet thickness as a function of cross-tail position Y. Black crosses denote individual events, and the red curve shows the median thickness in each Y bin, revealing systematically thicker current sheets on the dawnside. (b) Normalized thickness distributions for Type I (black) and Type II (gray) current sheets. Red asterisks indicate the ratio $N_{Type\ I}/N_{Type\ II}$ in each thickness bin (right axis), showing that thicker current sheets are more likely to remain quasi-laminar. (c) and (d) Dependence of inertial-range and kinetic-range spectral slopes, respectively, on current sheet thickness for Type II events. Thinner current sheets tend to exhibit shallower inertial-range slopes and steeper kinetic-range slopes, indicating more developed turbulence. Black crosses represent individual events, and red curves denote median values in each thickness bin. These results suggest a two-level picture. On a global scale, enhanced heavy-ion populations on the dusk side may suppress magnetic reconnection and turbulence, resulting in relatively stronger turbulent activity on the dawn side. On a local scale, however, within both the dawn side and dusk side, reconnection and turbulence preferentially occur in thinner current sheets, which are more susceptible to turbulent development.